\documentclass[prb,showpacs,twocolumn,floatfix,superscriptaddress]{revtex4}
\usepackage{amssymb}
\usepackage{makeidx}
\usepackage{amsmath}
\usepackage{graphicx}
\usepackage{amsfonts}

\begin{document}

\title{Resonant excitations of single and two-qubit systems coupled to a
tank circuit}
\author{S.N.~Shevchenko}
\affiliation{B.Verkin Institute for Low Temperature Physics and Engineering, 47 Lenin
Ave., 61103, Kharkov, Ukraine}
\author{S.H.W.~van~der~Ploeg}
\affiliation{Institute of Photonic Technology, P.O. Box 100239, D-07702 Jena, Germany}
\author{M.~Grajcar}
\affiliation{Institute of Photonic Technology, P.O. Box 100239, D-07702 Jena, Germany}
\affiliation{Department of Experimental Physics, Comenius University, SK-84248
Bratislava, Slovakia}
\author{E.~Il'ichev}
\affiliation{Institute of Photonic Technology, P.O. Box 100239, D-07702 Jena, Germany}
\author{A.N.~Omelyanchouk}
\affiliation{B.Verkin Institute for Low Temperature Physics and Engineering, 47 Lenin
Ave., 61103, Kharkov, Ukraine}
\author{H.-G.~Meyer}
\affiliation{Institute of Photonic Technology, P.O. Box 100239, D-07702 Jena, Germany}
\date{\today }

\begin{abstract}
The interaction of flux qubits with a low frequency tank circuit is studied.
It is shown that changes in the state of the interacting qubits influence
the effective inductance and resistance of the circuit, which is the essence
of the so-called impedance measurement technique. The multiphoton resonant
excitations in both single flux qubits and pairs of coupled flux qubits are
investigated. In particular, we compare our theoretical results with recent
spectroscopy measurements, Landau-Zener interferometry, and the multiphoton
fringes.
\end{abstract}

\maketitle

\section{Introduction}

Quantum properties of the atom-photon interaction have been mainly a subject
of quantum optics and atomic physics. For example, the interaction between
an atom and a photon was studied by experiments with a single atom in a
cavity, this topic represents a textbook subject of quantum electrodynamics.%
\cite{Rai} Recently a similar configuration has been realized for
superconducting quantum circuits \cite{Wal} where Josephson qubits play the
role of atoms, while a microwave resonator replaces the cavity. Usually for
such kind of experiments the photon energy is close to the level separation $%
\Delta /h$ of the atom or qubit. In contrast, in the experiments of Refs.~%
\onlinecite{ili,Ilichev04,born,Lupascu,multiphoton,Izmalkov08,Grajcar07},
driven qubits are coupled to a $LC$ tank circuit with a resonance frequency
well below the level splitting of the qubits. In particular, our study here
is motivated by the recent experiments \cite{Izmalkov08} where driven one or
two flux qubits systems were inductively coupled to a low frequency
superconducting tank circuit.

The flux, or persistent current, qubit is a superconducting ring with three
Josephson junctions.\cite{Mooij} The circuit is characterized by a two
dimensional potential which, for suitable qubit and external parameters,
exhibits two minima. If the applied magnetic flux is half a flux quantum,
these minima have equal energies. In the flux-neighborhood of this point the
degeneracy is lifted due to finite tunneling probability between minima.
Therefore, the circuit forms an effective two-level quantum system.

Strictly speaking a system of qubits coupled to a resonant circuit should be
treated quantum mechanically, as in Refs.~ %
\onlinecite{Smirnov,Greenberg05,Greenberg07,Grajcar07,shoen07}. However, due
to the weak coupling of the qubits to the classical circuit the
qubits-oscillator system can be treated semiclassically \cite%
{Zorin02,Krech02,Greenberg02a,Greenberg02b,QIMT}. In this paper we study the
impact of the qubits on the tank circuit in terms of the effective
inductance and resistance of the tank. We show that, in some limiting cases,
the analyzed equations can be simplified resulting in a more transparent
description of the behavior of the investigated qubit system.

In this paper we will address the measurement with the resonant tank circuit
in three different driving regimes of the qubits: in the ground state
(without driving field), and in the weakly and strongly driven regimes. Such
regimes are useful for controlling the state of the qubit with the driving
field. Particularly interesting is the strongly driven regime, where due to
the interference between different Landau-Zener tunnelling events the state
of the system quasi-periodically depends on both the DC bias and the AC
driving amplitude.\cite{ShIF,Sillanpaa, Oliver, ShO, Ashhab} Motivated by
our experimental paper\cite{Izmalkov08}, we study the measurement technique
in application to both single and coupled flux qubits. Systems of coupled
qubits have been studied previously\cite{2qbs} as well as spectroscopy with
switching current readout.\cite{Majer,spectr3qbs} The study of the dynamical
driving of coupled qubits is important to form an entangled state (in
particular, preparation of maximally entangled Bell states), and to perform
two-qubit operations, such as a CNOT gate.\cite{Liu, Plantenberg}
Particularly, we study the multiphoton resonances in a two-qubit four-level
system. This is analogous to the resonances studied in Ref.~\onlinecite{Yu}
where a multilevel system based on a single flux qubit including the upper
levels was considered.

The remainder of this paper is organized as follows. In Sec.~II we derive
equations which describe the influence of the qubits on the tank circuit in
terms of the effective inductance and resistance. Limiting cases are
considered in Sec.~III for a single flux qubit coupled to the tank circuit.
In Sec.~IV equations for the coupled qubits system are formulated.
Numerically calculated results are presented in Secs.~V and VI for single
and coupled qubits respectively. Experimental results for the multiphoton
resonances in single flux qubit are shown in Sec.~VII.

\section{Description of the measurement with tank circuit}

\begin{figure}[h]
\includegraphics[width=6cm]{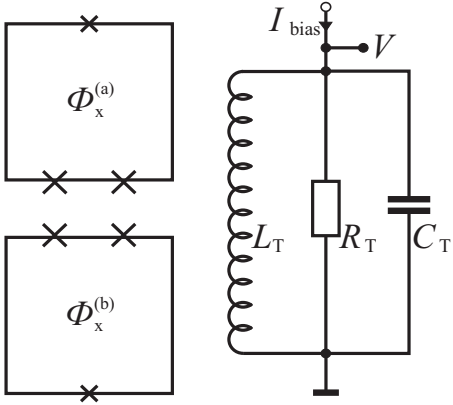}
\caption{Two-qubit system coupled to a tank circuit. The flux qubits are
pierced by magnetic fluxes $\Phi _{x}^{(a,b)}$ induced by the currents in
the controlling coils (not shown in the scheme) and by the current in the
tank's inductor. The qubits are coupled to each other and to the tank
circuit. The resonant tank circuit consists of the inductor $L_{T}$,
capacitor $C_{T}$, and resistor $R_{T}$; the circuit is biased with a RF
current $I_\mathrm{bias}$. The tank voltage $V$ is the measurable value.}
\label{scheme}
\end{figure}

Consider a tank circuit which consists of an inductor $L_{T}$, a capacitor $%
C_{T}$, and a resistor $R_{T}$ connected in parallel (see Fig. \ref{scheme}%
). The voltage $V$ in the current-biased tank circuit ($I_{\mathrm{bias}%
}=I_{A}\sin \omega _{\mathrm{rf}}t$), which is pierced by the external flux $%
\Phi _{e}$, is described by the following non-linear equation:\cite{QIMT}

\begin{equation}
C_{T}\overset{\centerdot \centerdot }{V}+\frac{\dot{V}}{R_{T}}+\frac{V}{L_{T}%
}=-\frac{\dot{\Phi}_{e}(V,\dot{V})}{L_{T}}+\dot{I}_{\mathrm{bias}}.
\label{eq_for_V}
\end{equation}%
The external flux $\Phi _{e}$ is assumed to be proportional to the coupling
parameter $k^{2}$ and to depend on time via voltage $V$ and its derivative $%
\dot{V}$.\cite{B-M}

The stationary oscillations in the non-linear system described by Eq. (\ref%
{eq_for_V}) can be reduced to oscillations in a linear system by making use
of the Krylov-Bogolyubov technique of asymptotic expansion.\cite{B-M}
Specifically, in the first order approximation with respect to the coupling
parameter $k^{2}$ and close to the principal resonance ($\omega_{\mathrm{rf}%
}\approx\omega_{T}\equiv1/\sqrt{L_{T}C_{T}}$) the equivalent linear system
is characterized by the effective resistance $R_{\mathrm{eff}}$ and
inductance $L_{\mathrm{eff}}$ as following:\cite{B-M}

\begin{equation}
C_{T}\overset{\centerdot\centerdot}{V}+\frac{\dot{V}}{R_{\mathrm{eff}}}+%
\frac{V}{L_{\mathrm{eff}}}=\dot{I}_{\mathrm{bias}},  \label{eff}
\end{equation}
\begin{equation}
V=v\cos(\omega_{\mathrm{rf}}t+\alpha),  \label{V}
\end{equation}

\begin{equation}
\frac{1}{R_{\mathrm{eff}}(v)}=\frac{1}{R_{T}}\left( 1-Q\beta_{s}(v)\right) ,
\label{Reff}
\end{equation}

\begin{equation}
\frac{1}{L_{\mathrm{eff}}(v)}=\frac{1}{L_{T}}\left( 1+\beta_{c}(v)\right) ,
\label{Leff}
\end{equation}
where $Q=\omega_{T}C_{T}R_{T}$ is the quality factor of the unloaded tank
circuit (at $\Phi_{e}=0$) and the functions $\beta_{s,c}(v)$ are given by:

\begin{equation}
\left\{
\begin{array}{c}
\beta _{s}(v) \\
\beta _{c}(v)%
\end{array}%
\right\} =\frac{1}{\pi v}\int\limits_{0}^{2\pi }\dot{\Phi}_{e}(v\cos \psi
,-v\omega _{\mathrm{rf}}\sin \psi )\left\{
\begin{array}{c}
\sin \psi \\
\cos \psi%
\end{array}%
\right\} d\psi .  \label{beta}
\end{equation}%
Let us also introduce notations $\varkappa _{s,c}$ for the partial
derivatives:
\begin{eqnarray}
\dot{\Phi}_{e}(V,\dot{V}) &=&\frac{\partial \Phi _{e}}{\partial V}\dot{V}+%
\frac{\partial \Phi _{e}}{\partial \dot{V}}\overset{\centerdot \centerdot }{V%
}  \notag \\
&\equiv &\varkappa _{s}(v,\psi )v\sin \psi +\varkappa _{c}(v,\psi )v\cos
\psi ,  \label{1}
\end{eqnarray}%
where $\psi =\omega _{\mathrm{rf}}t+\alpha $. If $\varkappa _{s,c}(v,\psi )$
are functions of either $\sin \psi $ or $\cos \psi $ only, i.e. $\varkappa
_{s,c}(v,\psi )=\varkappa _{s,c}^{\prime }(v,\sin \psi )$ or $\varkappa
_{s,c}(v,\psi )=\varkappa _{s,c}^{\prime }(v,\cos \psi )$, the Eq.~(\ref%
{beta}) leads to a simpler form for $\beta _{s,c}$:

\begin{align}
\beta _{s}(v)& =\frac{1}{\pi }\int\limits_{0}^{2\pi }\varkappa _{s}(v,\psi
)\sin ^{2}\psi d\psi ,  \label{2} \\
\beta _{c}(v)& =\frac{1}{\pi }\int\limits_{0}^{2\pi }\varkappa _{c}(v,\psi
)\cos ^{2}\psi d\psi .  \label{3}
\end{align}%
As a result of Eqs.~(\ref{eff}) and \ref{Leff} the resonant frequency $%
\omega _{\mathrm{eff}}$ becomes amplitude-dependent and is shifted by:

\begin{equation}
\frac{\omega _{\mathrm{eff}}(v)-\omega _{T}}{\omega _{T}}=\frac{1}{2}\beta
_{c}(v).  \label{weff}
\end{equation}%
The phase shift $\alpha$ and amplitude $v$ depend on the frequency detuning $%
\xi _{0}\equiv \frac{\omega _{T}-\omega _{\mathrm{rf}}}{\omega _{T}}$ and
the qubit state (via $\beta _{s,c}$). In the stationary regime they are
given by:

\begin{equation}
\left\{
\begin{array}{c}
v=I_{A}R_{\mathrm{eff}}\cos\alpha, \\
\tan\alpha=2Q\frac{R_{\mathrm{eff}}}{R_{T}}\left( \xi_{0}+\frac {L_{T}-L_{%
\mathrm{eff}}}{2L_{T}}\right) ,%
\end{array}
\right.  \label{with_L_R_eff}
\end{equation}
which can be rewritten in terms of the effective quality factor $Q_{\mathrm{%
eff}}$ and effective frequency shift $\xi_{\mathrm{eff}}$:

\begin{equation}
\left\{
\begin{array}{c}
v\sqrt{1+4Q_{\mathrm{eff}}^{2}\xi _{\mathrm{eff}}^{2}}=\frac{I_{A}Q_{\mathrm{%
eff}}}{\omega _{T}C_{T}}, \\
\tan \alpha =2Q_{\mathrm{eff}}\xi _{\mathrm{eff}}\text{ },%
\end{array}%
\right.  \label{with_Q_ksi_eff}
\end{equation}%
\begin{equation}
Q_{\mathrm{eff}}(v)=\omega _{T}C_{T}R_{\mathrm{eff}}(v)=\frac{Q}{1-Q\beta
_{s}(v)},
\end{equation}%
\begin{equation}
\xi _{\mathrm{eff}}(v)=\xi _{0}+\frac{1}{2}\beta _{c}(v)=\frac{\omega _{%
\mathrm{eff}}(v)-\omega _{\mathrm{rf}}}{\omega _{T}}.
\end{equation}%
Thus, the observable values -- the amplitude $v$ and the phase shift $\alpha
$ -- are defined by equations (\ref{with_L_R_eff}) or (\ref{with_Q_ksi_eff}%
), which depend on the response of the measurable system, $\Phi _{e}(V,\dot{V%
})$.

\section{Measurement of the persistent current qubit}

In this section we consider the system of the tank circuit coupled to a
persistent current qubit with geometrical inductance $L$ and average
persistent current $I_{\mathrm{qb}}$. The qubit is considered to be weakly
coupled to the tank circuit via a mutual inductance $M$. As we discussed
above, strictly speaking, the dynamics of the tank circuit has to be
considered jointly with the dynamics of the qubit. However, in this section
we consider two limiting cases, when the dynamics of the qubit can be
treated separately from the dynamics of the tank circuit. For simplification
we introduce phenomenologically the relaxation time $T_{1}$ (which can be
caused by the tank as well) and consider the weak coupling limit, $k^{2}=%
\frac{M^{2}}{LL_{T}}\ll 1$.

\subsection{Low-quality\ qubit ($T_{1}\ll T$): phase shift probes effective
inductance of qubit}

When all the qubit's characteristic times, and in particular the relaxation
time $T_{1}$, are smaller than the tank's period $T=2\pi /\omega _{T}$, the
equations can be simplified, since the equations for the tank voltage can be
averaged over the fast oscillating terms. This averaging is assumed
henceforth for all values in this subsection. Then the time derivative of
the flux $\Phi _{e}$, induced by the qubit in the tank circuit can be
described as:
\begin{equation}
\dot{\Phi}_{e}=M\dot{I}_{\mathrm{qb}}=M\frac{\partial I_{\mathrm{qb}}}{%
\partial \Phi }\dot{\Phi},
\end{equation}%
where $\Phi =\Phi _{\mathrm{dc}}+MI_{L}$ is the flux in the qubit's loop
(here\cite{QIMT} we ignore the small self-induced flux $-LI_{\mathrm{qb}}$),
which consists of the time-independent part $\Phi _{\mathrm{dc}}$ and of the
flux, induced by the current $I_{L}$ in the tank's inductor. This can be
rewritten by introducing the (inverse) effective inductance of the qubit, $%
\mathcal{L}^{-1}=\frac{\partial I_{\mathrm{qb}}(\Phi )}{\partial \Phi }$,
and the inductance value, which characterizes the response of the qubit, $%
\widetilde{L}=M^{2}\mathcal{L}^{-1}$. Then it follows:
\begin{equation}
\dot{\Phi}_{e}=\widetilde{L}(I_{L})\dot{I}_{L};  \label{Fi_e}
\end{equation}%
and for the tank voltage we have:\cite{QIMT}
\begin{equation}
V=L_{T}\dot{I}_{L}-\dot{\Phi}_{e}=(L_{T}-\widetilde{L}(I_{L}))\dot{I}_{L}.
\end{equation}%
Since $\widetilde{L}\propto k^{2}$ and $\widetilde{L}\ll L_{T}$, in the
first approximation in $k^{2}$ we can insert $I_{L}$ in the r.h.s. of Eq. (%
\ref{Fi_e}) found from the above equation:

\begin{equation}
I_{L}(t)\approx \frac{1}{L_{T}}\int Vdt=\frac{v}{\omega _{T}L_{T}}\sin
(\omega _{\mathrm{rf}}t+\alpha ).  \label{I_L}
\end{equation}%
Then by Eqs. (\ref{1}-\ref{3}) we have
\begin{equation}
\beta _{s}=0\text{, }\beta _{c}=\frac{1}{\pi }\int\limits_{0}^{2\pi }k^{2}L%
\mathcal{L}^{-1}(v,\psi )\cos ^{2}\psi d\psi ,  \label{beta_for_quick}
\end{equation}%
where the qubit's effective inductance is defined by the total flux $\Phi $,
piercing the qubit's loop:%
\begin{equation}
\mathcal{L}^{-1}(v,\psi )\equiv \left. \frac{\partial I_{\mathrm{qb}}(\Phi )%
}{\partial \Phi }\right\vert _{\Phi =\Phi _{\mathrm{dc}}+\frac{M}{%
L_{T}\omega _{T}}v\sin \psi }.  \label{Jos_induct}
\end{equation}%
It follows:%
\begin{equation}
\tan \alpha \approx 2Q\xi _{0}+Q\beta _{c},\text{ \ }v\approx I_{A}R_{T}\cos
\alpha .  \label{a_and_v}
\end{equation}%
Actually, this is a generalization of the result of Ref.~%
\onlinecite{Greenberg02a} for the case when the qubit can be in the excited
state (which is taken into account by the expectation value of the current $%
I_{\mathrm{qb}}$).

If the bias current amplitude $I_{A}$ is small enough to be ignored in Eq.~(%
\ref{Jos_induct}), where $v\sim I_{A}R_{T}$, then:\cite{QIMT}
\begin{eqnarray}
\tan\alpha&\approx&2Q\xi_{0}+k^{2}Q\frac{L}{\mathcal{L}},\text{ \ }v\approx
I_{A}R_{T}\cos\alpha,  \notag \\
\mathcal{L}^{-1}&\approx&\frac{\partial I_{\mathrm{qb}}(\Phi_{\mathrm{dc}})%
} {\partial\Phi_{\mathrm{dc}}},  \label{QIMT}
\end{eqnarray}
which means that at the resonant frequency ($\xi_{0}=0$) the tank's phase
shift $\alpha$\ is approximately proportional to the inverse inductance of
the qubit $\mathcal{L}^{-1}$.

\subsection{Higher-quality\ qubit ($T_{1}\lesssim T$): effective resistance
due to qubit's lagging}

Consider the case when the qubit relaxation time $T_{1}$ is of the same
order as the tank's period $T$ (namely, $T_{1}\lesssim T$). We assume
exponential delayed response of the qubit $1-\exp (-t/T_{1})$ which can be
phenomenologically described as a qubit response lagging relative to the
tank circuit (as e.g. in Ref.~\onlinecite{Metzger}). Therefore, Eq.~(\ref%
{Fi_e}) is replaced by:
\begin{equation}
\dot{\Phi}_{e}(t)=\widetilde{L}(I_{L}(t^{\prime }))\dot{I}_{L}(t^{\prime });
\label{Fi_e_}
\end{equation}%
where $t^{\prime }=t-T_{1}$ is retarded time. Thus, the qubit's response
depends on the current in the tank $I_{L}=I_{L}(t^{\prime })$, which is
given by:
\begin{eqnarray}
I_{L}(t^{\prime }) &\approx &\frac{v}{\omega _{T}L_{T}}\sin (\omega _{%
\mathrm{rf}}t^{\prime }+\alpha )  \notag \\
&=&\frac{v}{\omega _{T}L_{T}}\left( C\sin (\omega _{\mathrm{rf}}t+\alpha
)-S\cos (\omega _{\mathrm{rf}}t+\alpha )\right) ,  \notag \\
&&  \label{prime}
\end{eqnarray}%
where $S=\sin (\omega _{\mathrm{rf}}T_{1})$ and $C=\cos (\omega _{\mathrm{rf}%
}T_{1})$. For the sake of simplicity we consider small bias current
approximation, in this case Eqs. (\ref{Fi_e_}-\ref{prime}) and definitions (%
\ref{1}-\ref{3}) result in: $\beta _{s}\approx k^{2}\frac{L}{\mathcal{L}}S$
and $\beta _{c}\approx k^{2}\frac{L}{\mathcal{L}}C$. Then, from Eqs.~(\ref%
{Reff}, \ref{Leff}) one gets:
\begin{align}
\frac{L_{T}}{L_{\mathrm{eff}}}& \approx 1+C\cdot k^{2}Q\frac{L}{\mathcal{L}},
\\
\frac{R_{T}}{R_{\mathrm{eff}}}& \approx 1-S\cdot k^{2}Q\frac{L}{\mathcal{L}},
\end{align}%
\begin{align}
\tan \alpha & \approx \frac{2Q\xi _{0}+C\cdot k^{2}QL/\mathcal{L}}{1-S\cdot
k^{2}QL/\mathcal{L}},  \label{a_with_S} \\
\frac{v}{I_{A}R_{T}}& \approx \frac{\cos \alpha }{1-S\cdot k^{2}QL/\mathcal{L%
}}.  \label{V_with_S}
\end{align}%
Consider these expressions in the first approximation in $k^{2}QL/\mathcal{L}
$; for $\xi _{0}=0$ we obtain the following important result:%
\begin{align}
\tan \alpha & \approx C\cdot k^{2}QL/\mathcal{L}, \\
\frac{v}{I_{A}R_{T}}& \approx 1+S\cdot k^{2}QL/\mathcal{L},  \notag
\end{align}%
which, in particular, shows that as $C\rightarrow 0$ there can be changes in
the amplitude $v$ without changes in the phase shift $\alpha $ as well as in
shift of the resonant frequency (Eq.~(\ref{weff})). It is important to note
that both the phase shift and amplitude are related to the qubit's effective
inductance $\mathcal{L}$, which explains their similar behavior in
experiment. These equations might be useful for qualitative analysis of
experimental results.

\section{Coupled flux qubits}

\subsection{Hamiltonian}

Close to its degeneracy point the flux qubit\cite{Mooij} can be described by
the pseudospin Hamiltonian:

\begin{equation}
H_{\mathrm{1qb}}=-\frac{\Delta }{2}\sigma _{1}-\frac{\epsilon (t)}{2}\sigma
_{3},  \label{H1q}
\end{equation}%
where the diagonal term $\epsilon $ is the energy bias, the off-diagonal
term $\Delta $ is the tunneling amplitude between the wells (which
corresponds to the definite directions of the current in the loop) and $%
\sigma _{j}$ are Pauli matrices.

For the system of coupled qubits the effective Hamiltonian is:
\begin{equation}
H_{\mathrm{2qbs}}=\sum\limits_{i=1,2}\left( -\frac{\Delta _{i}}{2}\sigma
_{1}^{(i)}-\frac{\epsilon _{i}(t)}{2}\sigma _{3}^{(i)}\right) +\frac{J}{2}%
\sigma _{3}^{(1)}\sigma _{3}^{(2)},  \label{H}
\end{equation}%
where $J$ is the coupling energy between qubits, and $\sigma _{1}^{(i)}$, $%
\sigma _{3}^{(i)}$ are the Pauli matrices in the basis $\{\left\vert
\downarrow \right\rangle ,\left\vert \uparrow \right\rangle \}$ of the
current operator in the $i$-th qubit; namely, $\sigma _{a}^{(1)}=\sigma
_{a}\otimes \sigma _{0}$, $\sigma _{a}^{(2)}=\sigma _{0}\otimes \sigma _{a}$%
, $\sigma _{0}$ is the unity matrix. The current operator is given by: $%
I_{i}=-I_{p}^{(i)}\sigma _{3}^{(i)},$ with $I_{p}^{(i)}$ the absolute value
of the persistent current in the $i$-th qubit; then the eigenstates of $%
\sigma _{3}$\ correspond to the clockwise ($\sigma _{3}\left\vert \downarrow
\right\rangle =-\left\vert \downarrow \right\rangle $) and counterclockwise (%
$\sigma _{3}\left\vert \uparrow \right\rangle =\left\vert \uparrow
\right\rangle $) current in the $i$-th qubit. The tunneling amplitudes $%
\Delta _{i}$ are assumed to be constants. The biases $\epsilon
_{i}=2I_{p}^{(i)}\Phi _{0}f^{(i)}(t)$ are controlled by the dimensionless
magnetic fluxes $f^{(i)}(t)=\Phi _{i}/\Phi _{0}-1/2$ through $i$-th qubit.
These fluxes consist of three components:
\begin{equation}
f^{(i)}(t)=f_{i}+\frac{M_{i}I_{L}}{\Phi _{0}}+f_{\mathrm{ac}}\sin \omega t.
\label{fx}
\end{equation}%
Here $f_{i}$ is the adiabatically changing magnetic flux, experimentally
applied by the coil and additional DC lines. The second term describes the
flux induced by the current $I_{L}$ in the tank coil, to which the $i$-th
qubit is coupled with the mutual inductance $M_{i}$. And $f_{\mathrm{ac}%
}\sin \omega t$ is the harmonic time-dependent component driving the qubit,
typically applied by an on-chip microwave antenna.

\subsection{Entanglement}

It is convenient to present the density matrix for two qubits in the
following form:
\begin{eqnarray}
\rho &=&\frac{R_{\alpha \beta }}{4}\sigma _{\alpha }\otimes \sigma _{\beta }
\notag \\
&=&\frac{R_{00}}{4}\sigma _{0}\otimes \sigma _{0}+\frac{R_{a0}}{4}\sigma
_{a}\otimes \sigma _{0}  \notag \\
&+&\frac{R_{0b}}{4}\sigma _{0}\otimes \sigma _{b}+\frac{R_{ab}}{4}\sigma
_{a}\otimes \sigma _{b},  \label{dens_mat}
\end{eqnarray}%
which was shown to be suitable for both the definition and the calculation
of the entanglement and other characteristics in multi-qubit system.\cite%
{Schlienz,Liu,Love,Ivanchenko} Here $0\leq \left\{ \alpha ,\beta \right\}
\leq 3$ and $1\leq \left\{ a,b\right\} \leq 3$; the summation over twice
repeating indices is assumed. The two vectors $R_{a0}$ and $R_{0b}$,
so-called coherence vectors or Bloch vectors, determine the properties of
the individual qubits, while the tensor $R_{ab}$ (the correlation tensor)
accounts for the correlations.\cite{Schlienz} (In the notations of Ref.~%
\onlinecite{Schlienz}: $\lambda _{a}(1)=R_{a0}$, $\lambda _{b}(2)=R_{0b}$, $%
K_{ab}=R_{ab}$.) Following Ref.~\onlinecite{Schlienz}, we choose the measure
of entanglement $\mathcal{E}$ to be the following:
\begin{equation}
\mathcal{E}=\frac{1}{3}\mathrm{Tr}\left( M^{T}M\right) ,\text{ }%
M_{ab}=R_{ab}-R_{a0}R_{0b}.  \label{E}
\end{equation}%
This measure of entanglement fulfills certain reasonable requirements ($%
0\leq \mathcal{E}\leq 1$), in particular, $\mathcal{E}=0$ for any product
state and $\mathcal{E}=1$ for any pure state with vanishing Bloch vectors $%
R_{a0}$ and $R_{0b}$, corresponding to maximum entangled states (see Ref.~%
\onlinecite{Schlienz} for more detail).

\subsection{Liouville equation}

The dynamics of the density matrix without taking into account the
relaxation processes can be described by the Liouville equation: $i\dot{\rho}%
=\left[ H,\rho \right] $, which is generally speaking a complex equation. We
set both $\hbar =1$\ and $k_{B}=1$ throughout. By the proper choice of the
parametrization the Liouville equation can be written in the form of a
system with a minimal number of real equations ($3$ -- for one qubit and $15$
for two qubits).

To deal with the Liouville equation, we make use of the parametrization
(decomposition) of the density matrix as described by Eq.~(\ref{dens_mat}).
This allows to benefit from the properties of the density matrix, namely
from its hermiticity (then $R_{\alpha \beta }$ are real numbers) and from
the normalization condition, $\mathrm{Tr}\rho =1$ (then $R_{00}=1$). It
follows that the density matrix is parameterized by $15$ independent real
values. It is useful to note that:
\begin{equation}
\mathrm{Tr}\left( \rho \sigma _{a}^{(1)}\right) =R_{a0},\text{ }\mathrm{Tr}%
\left( \rho \sigma _{b}^{(2)}\right) =R_{0b}.  \label{Ra0}
\end{equation}%
After straightforward algebra the Liouville equation yields:%
\begin{eqnarray}
\dot{R}_{i0} &=&\varepsilon _{mni}B_{m}^{(1)}R_{n0}+\varepsilon
_{3ni}JR_{n3},  \notag \\
\dot{R}_{0j} &=&\varepsilon _{mnj}B_{m}^{(2)}R_{0n}+\varepsilon
_{3nj}JR_{3n},  \notag \\
\dot{R}_{ij} &=&\varepsilon _{mni}B_{m}^{(1)}R_{nj}+\varepsilon
_{mnj}B_{m}^{(2)}R_{in}  \notag \\
&+&\delta _{j3}\varepsilon _{3ni}JR_{n0}+\delta _{i3}\varepsilon
_{3nj}JR_{0n}.  \label{sys_15_eqs}
\end{eqnarray}%
where $\mathbf{B}^{(i)}$ are \textquotedblleft local magnetic
fields\textquotedblright , which for the flux qubits are defined as $\mathbf{%
B}^{(i)}=\left( -\Delta _{i},0,-\epsilon _{i}\right) $, and $\varepsilon
_{mni}$ is the Levi-Civita symbol.

\subsection{Effective inductance of coupled qubits}

For describing the effective inductance of coupled qubits it is important to
note that the current in the $i$-th qubit depends on the fluxes in both
qubits, $I_{\mathrm{qb}}^{(i)}=I_{\mathrm{qb}}^{(i)}(\Phi _{x}^{(a)},\Phi
_{x}^{(b)})$. Here the fluxes $\Phi _{x}^{(a,b)}$ consist of a DC part, $%
\Phi _{i}$, and of the flux generated by the current in the tank coil, $%
I_{L} $: $\Phi _{x}^{(i)}=\Phi _{i}+MI_{L}$. For simplicity we assume here
that the qubit-tank mutual inductance $M$ is the same for both qubits (for
more detail see Ref.~\onlinecite{QIMT}). Then the time derivative is:%
\begin{equation}
\dot{I}_{\mathrm{qb}}^{(i)}=\left( \frac{\partial }{\partial \Phi _{x}^{(a)}}%
+\frac{\partial }{\partial \Phi _{x}^{(b)}}\right) I_{\mathrm{qb}%
}^{(i)}\cdot M\dot{I}_{L}.
\end{equation}%
In the limit of small bias current in the tank we can substitute $\Phi _{i}$
for $\Phi _{x}^{(i)}$ and define the qubit effective inductance with the
symmetric change of the flux bias in both qubits\cite{Grajcar05} $\mathbf{%
\Phi }_{x}=(\Phi _{a},\Phi _{b})$:%
\begin{equation}
\mathcal{L}_{i}^{-1}=\frac{\partial }{\partial \mathbf{\Phi }_{x}}I_{\mathrm{%
qb}}^{(i)}\equiv \left( \frac{\partial }{\partial \Phi _{a}}+\frac{\partial
}{\partial \Phi _{b}}\right) I_{\mathrm{qb}}^{(i)}.  \label{symm_deriv}
\end{equation}%
Then for the case of low-quality~qubits, when their characteristic times are
smaller than the tank's period, analogously to Sec.~III.A, we obtain at the
resonance frequency ($\xi _{0}=0$):%
\begin{equation}
\tan \alpha \approx \sum \Xi _{i}\frac{\Phi _{0}}{I_{p}^{(i)}}\mathcal{L}%
_{i}^{-1},  \label{alpha_4_2}
\end{equation}%
where we introduced the notation%
\begin{equation}
\Xi _{i}=k^{2}Q\frac{L_{i}I_{p}^{(i)}}{\Phi _{0}}.
\end{equation}

\section{Results for single flux qubit}

\subsection{Spectroscopy}

In Ref. \onlinecite{Izmalkov08} it was shown that the qubit parameters can
be determined both by measurements in the ground state or by employing a
spectroscopic measurements when the qubit is resonantly excited. In this
section we show related numerically calculated graphs, making use of the
results of previous sections.

Consider a qubit biased with a DC flux $\Phi _{\mathrm{dc}}$ and driven with
an AC flux $\Phi _{\mathrm{ac}}\sin \omega t$, introducing
\begin{equation*}
f_{\mathrm{dc}}=\Phi _{\mathrm{dc}}/\Phi _{0}-1/2\text{ and }f_{\mathrm{ac}%
}=\Phi _{\mathrm{ac}}/\Phi _{0}.
\end{equation*}%
In order to get the effective inductance $\mathcal{L}$, as defined by Eq. (%
\ref{Jos_induct}), we have to calculate the average current in qubit: $I_{%
\mathrm{qb}}=\left\langle I\right\rangle =\mathrm{Tr}\left( \rho I\right) $,
where $I=-I_{p}\sigma _{3}$ is the current operator defined with the
amplitude $I_{p}$ and the Pauli matrix $\sigma _{3}$. We calculate the
reduced density matrix $\rho $ with the Bloch equations,\cite{Blum, ShKOK,
multiphoton} which include phenomenological relaxation times, $T_{1}$ and $%
T_{2}$. It is convenient to express the density matrix in the energy
representation: $\rho =\left( 1/2\right) \left( \tau_{0}+X\tau _{1}+Y\tau
_{2}+Z\tau _{3}\right) $, where $\tau _{i}$ are the Pauli matrices for this
basis and $\tau_0$ stands for the unity matrix. $Z$ is equal to the
difference between the populations of the ground and excited states. As a
result the effective inductance is given by:\cite{QIMT}
\begin{equation}
\mathcal{L}^{-1}=I_{p}\frac{\partial }{\partial \Phi }\left\{ \frac{\Delta }{%
\Delta E}X-\frac{2I_{p}\Phi }{\Delta E}Z\right\} ,  \label{inverse_induct}
\end{equation}%
where $\Delta E=\sqrt{\Delta ^{2}+\left( 2I_{p}\Phi \right) ^{2}}$.

\begin{figure}[h]
\includegraphics[width=9cm]{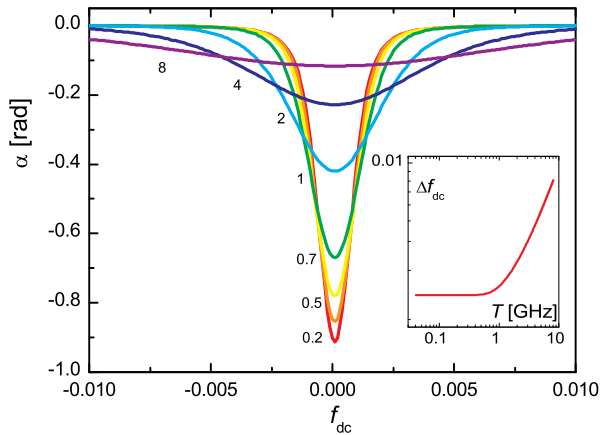}
\caption{(Color online). Influence of temperature on the ground state
measurement: the dependence of the tank phase shift on the flux detuning $f_{%
\mathrm{dc}}=\Phi _{\mathrm{dc}}/\Phi _{0}-1/2$, when the qubit is thermally
excited. The numbers next to the curves stand for the temperature in GHz.
Inset: temperature dependence of the width $\Delta f_{\mathrm{dc}}$ of the
dip at half-depth in the phase shift, shown in the main panel.}
\label{vs_T}
\end{figure}

First consider the ground-state measurement, which is described by Eqs.~(\ref%
{beta_for_quick}-\ref{a_and_v}). These equations not only allow us to
reproduce the results of Refs.~\onlinecite{Greenberg02a} and %
\onlinecite{Grajcar04}, valid for the case where the system is in the ground
state, but also describe the situation when the qubit is excited. Consider
the influence of temperature, when the qubit is in a thermal mixture of the
ground and excited states. In this case $X=0$ and $Z=\tanh (\Delta E/2T)$.
The resulting tank phase shift is shown in Fig. \ref{vs_T} for the following
parameters:\cite{Grajcar04} $\Delta /h=2\cdot 0.65$\ GHz, $I_{p}\Phi
_{0}/h=930$\ GHz, $\omega _{T}/2\pi =32.675$\ MHz, $LI_{p}/\Phi _{0}=0.0055$%
, $M/L=0.725$, $Q=725$, $k=0.02$. The accurate account of $Z$\ allows us to
describe the widening of the phase shift dip, as shown in the inset in Fig.~%
\ref{vs_T}, which was reported in Ref.~\onlinecite{Grajcar04}. The widening
is due to the term that comes from differentiating the $\tanh $ in Eq.~(\ref%
{inverse_induct}); this term becomes relevant for temperatures larger than $%
\Delta =1.3$\ GHz, and results in the exponential rise of the width for $%
T>T^{\ast }=\Delta $\ ($\frac{d}{dx}\tanh x\simeq 4\exp (-2x)$\ at $x>1$).

\begin{figure}[h]
\includegraphics[width=8cm]{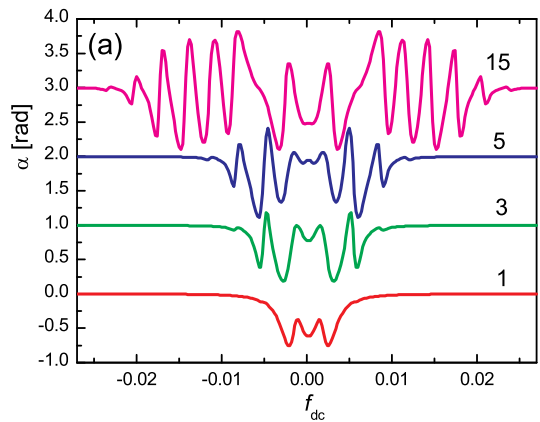} \includegraphics[width=8cm]{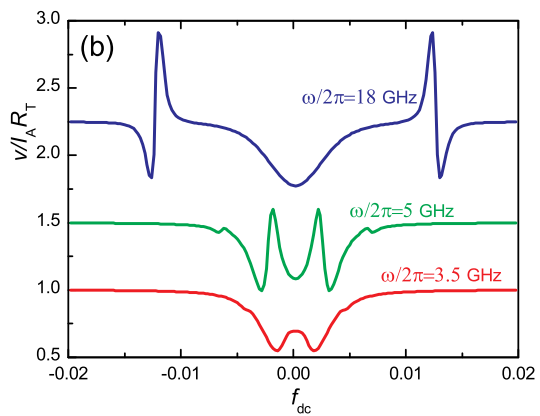}
\caption{(Color online). Curves for a resonantly excited flux qubit: (a) the
phase shift and (b) the amplitude of the tank voltage versus flux detuning $%
f_{\mathrm{dc}}$. In panel (a) the numbers next to the curves stand for the
driving amplitude $f_{\mathrm{ac}}$ multiplied by $10^{3}$; in panel (b) the
curves are for the amplitudes $f_{\mathrm{ac}}\cdot 10^{3}=1,1.5,3$ from
bottom to top.}
\label{spectroscopy}
\end{figure}

Now consider the spectroscopical measurement, where the qubit is driven with
the AC flux. In Fig.~\ref{spectroscopy}(a) we demonstrate the dependence of
the phase shift $\alpha $ on the bias flux $f_{\mathrm{dc}}$ at $\omega _{%
\mathrm{rf}}=\omega _{T}$ for different driving amplitudes at the driving
frequency $\omega /2\pi =4.15$ GHz. The parameters for plotting the graph
were taken as in the related experiment\cite{Izmalkov08} and are given in
Table~\ref{tab:table1}.
\begin{table}[tbp]
\caption{The parameters used for plotting Fig.~\protect\ref{spectroscopy}. $%
\Delta $, $I_{p}\Phi _{0}$, $\Gamma _{1}$, and $\Gamma _{2}$ are tunneling
amplitude, energy bias, relaxation rate, and dephasing rate, respectively.
The $\Xi $ describes the coupling between qubit and tank circuit. All
parameters are in units of $h\cdot $GHz except $\Xi $ which is
dimensionless. }
\label{tab:table1}%
\begin{ruledtabular}
\begin{tabular}{cccccc}
    $\Delta$&$I_p\Phi_0$&$T$&$\Gamma_{1}$&$\Gamma_{2}$
    &$\Xi$\\
\hline
3.5 & 700 & 1.4 & 0.7 & 0.7 & $2.6\cdot10^{-3}$\\
\end{tabular}
\end{ruledtabular}
\end{table}
In Fig.~\ref{spectroscopy} the upper curves are shifted vertically for
clarity. In Fig.~\ref{spectroscopy}(b) we plot the amplitude $v$ versus the
bias flux $f_{\mathrm{dc}}$ with the phenomenological lagging parameter $%
S=0.8$ for several values of the driving frequency $\omega $, making use of
Eqs.~(\ref{a_with_S}-\ref{V_with_S}). In the experimental case the positions
of these resonances at a given driving frequency allow to determine the
energy structure of the measured qubit.\cite{Izmalkov08}
Figure \ref{spectroscopy} demonstrates the effect described in section III:
for $S\neq 0$ both the phase shift $\alpha $ and the amplitude $v$\ depend
on $\mathcal{L}^{-1}$, which results in the alternation of peak and dip
around the location of the resonances. We also note that, in addition, the
terms nonlinear in $S$ distort this structure. The second-order term for
example, is proportional to $\mathcal{L}^{-2}(3S^{2}-1)$. For values $S>1/%
\sqrt{3}$ this term leads to increasing of the peak and decreasing of the
dip leading to the asymmetry visible in the upper curve of Fig.~\ref%
{spectroscopy}(b).

\begin{figure}[h]
\includegraphics[width=8cm]{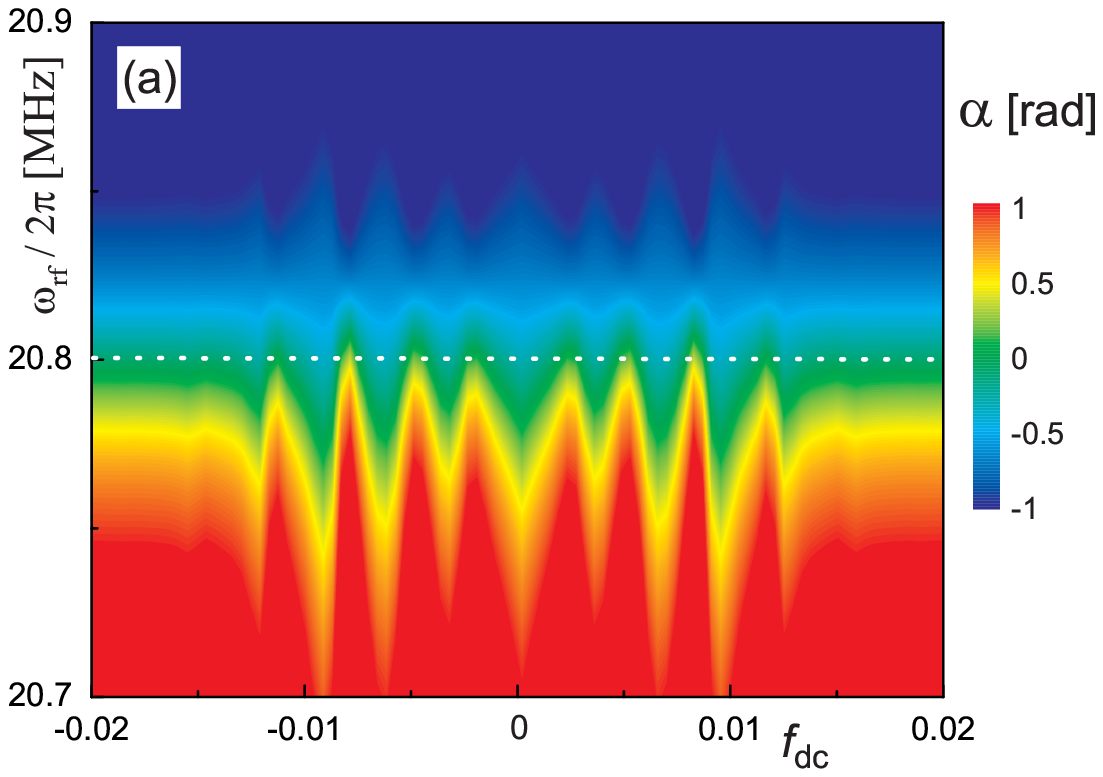} \includegraphics[width=8cm]{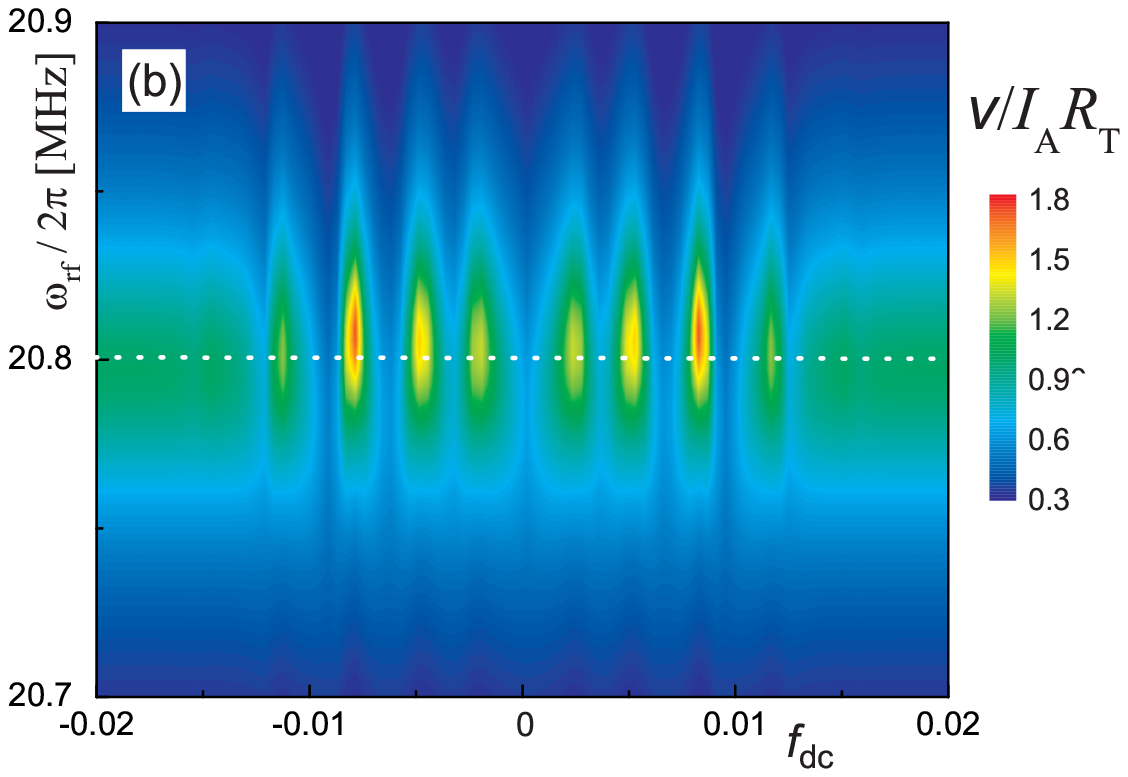}
\caption{(Color online). Multi-photon excitations of a single flux qubit.
(a) dependence of the phase shift $\protect\alpha $ and (b) the amplitude $v$
on the bias current frequency $\protect\omega _{\mathrm{rf}}$ and the flux
detuning $f_{\mathrm{dc}}$; parameters: $f_{\mathrm{ac}}=8\cdot 10^{-3}$, $%
\protect\omega /2\protect\pi =4.15$\ GHz, $S=0.8$.}
\label{multiph_1qb}
\end{figure}

In Fig.~\ref{multiph_1qb} we plot the phase shift $\alpha $ and the
amplitude $v$ as functions of the bias current frequency $\omega _{\mathrm{rf%
}}$ and the flux detuning $f_{\mathrm{dc}}$ with the phenomenological
lagging parameter $S$ for the flux qubit with the parameters given in Table.~%
\ref{tab:table1}. The dashed white line shows the tank resonance frequency $%
\omega _{\mathrm{rf}}/2\pi =\omega _{T}/2\pi =20.8$ MHz. Note that for the
lagging parameter close to $1$ ($S=0.8$) the changes in the phase shift in
Fig.~\ref{multiph_1qb}(a) are small at the resonance frequency (along the
white line) while the voltage amplitude in Fig.~\ref{multiph_1qb}(b) changes
substantially. Such changes of the tank effective resistance or,
equivalently, quality factor were studied in Ref.~\onlinecite{Grajcar07}.

\subsection{Landau-Zener interferometry}

\begin{figure}[h]
\includegraphics[width=8cm]{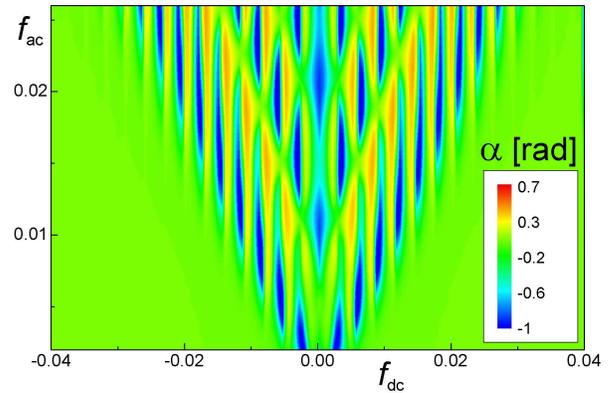}
\caption{(Color online). Landau-Zener interferometry: dependence of the tank
phase shift on the flux detuning $f_{\mathrm{dc}}=\Phi_{\mathrm{dc}%
}/\Phi_{0}-1/2$ and on the driving flux amplitude $f_{\mathrm{ac}%
}=\Phi_{ac}/\Phi_{0}$.}
\label{LZI}
\end{figure}

LZ interferometry is demonstrated in Fig.~\ref{LZI} as the dependence of the
tank voltage phase shift $\alpha$ on the microwave amplitude $f_{\mathrm{ac}%
} $ and the DC flux bias $f_{\mathrm{dc}}$. The qubit parameters were taken
as for Fig.~\ref{spectroscopy} (Table~\ref{tab:table1}) and $%
\omega/2\pi=4.15 $\ GHz.

The multiphoton resonances at discrete DC bias $f_{\mathrm{dc}}$\ (which
controls the distance between energy levels) are clearly visible. These
resonances appear when the energy of $n$ photons matches the qubit's energy
levels: $n\cdot \hbar \omega =\Delta E(f_{\mathrm{dc}})$. Note also the
quasi-periodical character of the dependence on the AC flux amplitude $f_{%
\mathrm{ac}}$, this is known as St\"{u}ckelberg oscillations; the comparison
of such graph to the experimental analogue \cite{Izmalkov08} (namely the
estimation of the period of St\"{u}ckelberg oscillations) allows the
relation of the microwave power to the AC flux amplitude $f_{\mathrm{ac}}$
to be determined.

\subsection{Impact of finite bias current}

\begin{figure}[h]
\includegraphics[width=8cm]{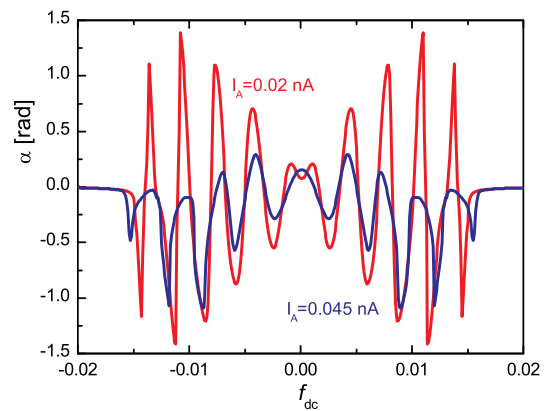}
\caption{(Color online). Impact of the finite bias current: dependence of
the phase shift $\protect\alpha$ on the bias flux $f_{\mathrm{dc}}$ for
different values of the bias current amplitude $I_{A}$, for $\protect\omega/2%
\protect\pi=4$ GHz, $f_{\mathrm{ac}}=8\cdot10^{-3}$.}
\label{impact}
\end{figure}

Consider the impact of the finite bias current on the tank's response. When
the bias current is small, its influence can be neglected and there is a
peak-and-dip structure, around the point where the qubit is resonantly
excited to the upper state, see Eq.~(\ref{QIMT}); these structures are
distorted by the non-linear terms when the current is increased. This is
demonstrated in Fig.~\ref{impact}, where we plot the dependence of the phase
shift $\alpha$ on the bias flux $f_{\mathrm{dc}}$ for two different values
of the bias current amplitude $I_{A}$. The qubit parameters were taken as
for Fig.~\ref{spectroscopy} (Table.~\ref{tab:table1}).

\textit{\ }

\section{Results for the two-qubit system}

\subsection{Resonant excitation}

\begin{widetext}

\begin{figure}[h]
\includegraphics[width=13cm]{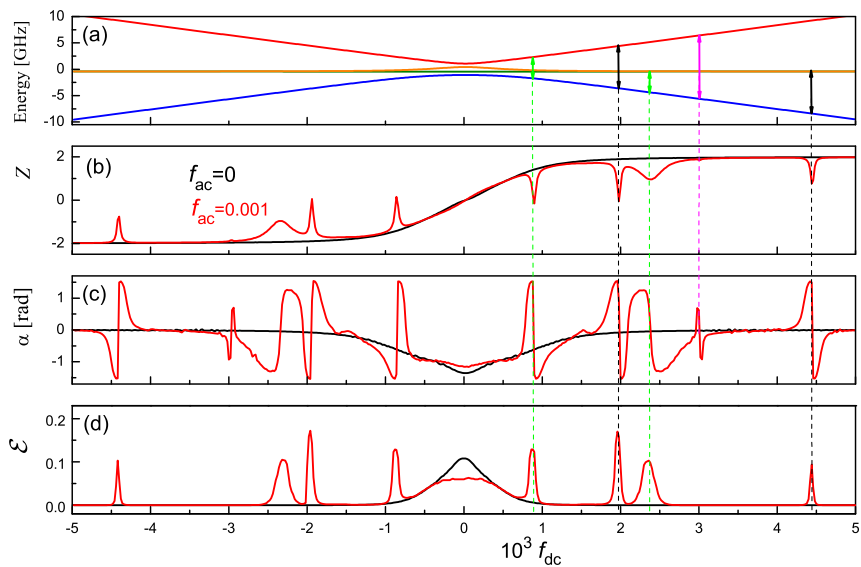}\caption{(Color online).
Multiphoton excitations of coupled flux qubits: (a) energy levels,
(b) total probability of the currents in qubits to flow clockwise
$Z$, (c) the tank phase shift $\alpha$, (d) the entanglement
measure $\mathcal{E}$ -- versus the bias
$f_{\mathrm{dc}}=f_{a}=f_{b}$.} \label{multiphoton}
\end{figure}

\end{widetext}

In this section we investigate the resonant excitation of a system of two
coupled flux qubits. First we will describe the effects of resonant
excitation in a system of two qubits in terms of its energy structure,
entanglement measure, and the observable tank circuit phase shift. Then we
will study the one-photon excitation of the system, which is used for the
spectroscopic measurements. Finally we will demonstrate the multiphoton
fringes. For calculations we will use the two sets of parameters as in Refs.~%
\onlinecite{2qbs} (i) and \onlinecite{Izmalkov08} (ii) which are given in
Table~\ref{tab:table2}.
\begin{table}[tbp]
\caption{The parameters of the two-qubit systems. All parameters are in
units of $h\cdot $GHz except $\Xi _{a,b}$ which are dimensionless.}
\label{tab:table2}%
\begin{ruledtabular}
\begin{tabular}{cccccccc}
$\ \ \ \ $&$\Delta_a$&$\Delta_b$&$I^{(a)}_p\Phi_0$&$I^{(b)}_p\Phi_0$&$J$&$\Xi_a$&$\Xi_b$\\
\hline
(i)  & 1.1 &  0.9 & 990 & 990 & 0.84 & $1.8\cdot10^{-3}$ & $1.8\cdot10^{-3}$\\
(ii) & 15.8 & 3.5 & 375 & 700 & 3.80 & $1.4\cdot10^{-3}$ & $2.6\cdot10^{-3}$\\
\end{tabular}
\end{ruledtabular}
\end{table}
Here, for clarity, we consider the case when the characteristic measurement
time $T$ is larger than the characteristic times of the dynamics of the
qubit ($T_{1}$). Then the tank circuit actually probes the incoherent
mixture of qubit's states and the time-averaged values of phase shift and
entanglement should be considered. We calculate the energy levels (by
diagonalizing the stationary Hamiltonian), the density matrix $\rho $, the
observable tank circuit phase shift $\alpha $ (which is defined with the
effective inductance of the qubits), and the entanglement measure $\mathcal{E%
}$\ by making use of equations (\ref{E}-\ref{alpha_4_2}). Then we plot Fig.~%
\ref{multiphoton} for the set of parameters (i) and the driving frequency $%
\omega /2\pi =4$\ GHz, assuming the symmetrical change of the DC flux: $%
f_{a}=f_{b}\equiv f_{\mathrm{dc}}$. Four energy levels are plotted in Fig.~%
\ref{multiphoton}(a). When the energy of $n$ photons ($n\cdot \hbar \omega $%
) matches the energy difference between any two levels $E_{j}$ and $E_{i}$,
the resonant excitation to the upper level is expected. Respectively, with
the green (gray), black, and magenta (dark gray) arrows of the length $4$, $8
$, and $12$\ GHz we show the places of possible one-, two-, and three-photon
excitations. The time-averaged total probability of the currents in two
qubits to flow clockwise, $Z=R_{03}+R_{30}$, is shown in Fig.~\ref%
{multiphoton}(b) to experience resonant excitation; black and red (dark
gray) lines correspond to $f_{\mathrm{ac}}=0$\ and $10^{-3}$. The resonances
appear as hyperbolic-like structures in the phase shift dependence in Fig. %
\ref{multiphoton}(c). The time-averaged entanglement measure $\mathcal{E}$\
has a peak in resonance, Fig.~\ref{multiphoton}(d). The entanglement measure
in a resonance increases due to the resonant formation of the superposition
of states; this provides a tool for resonantly controlling the entanglement
and on the other hand comparing Figs.~\ref{multiphoton}(c) and \ref%
{multiphoton}(d) is a method to probe the entanglement.

\subsection{Spectroscopy (one-photon excitation)}

\begin{figure}[h]
\includegraphics[width=8cm]{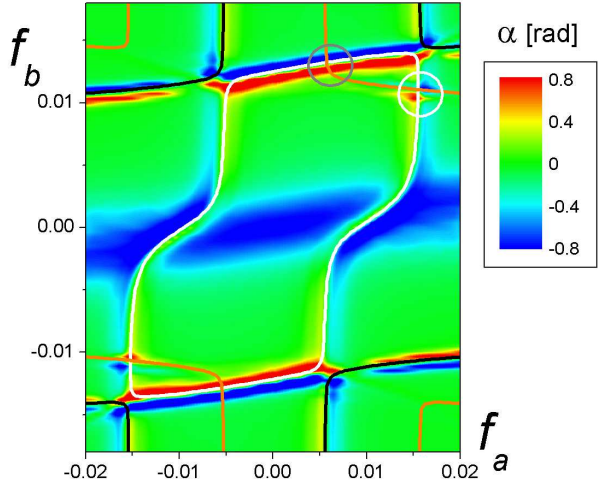}
\caption{(Color online). Resonant excitation of two coupled flux qubits: the
tank phase shift versus partial bias fluxes in two qubits, $f_{a}$ and $%
f_{b} $.}
\label{a_vs_fa_and_fb}
\end{figure}

A weakly driven system of qubits can be resonantly excited to an upper level
when the driving frequency matches the distance between the energy levels.
This can be used for the spectroscopic measurement of the energy structure
of the system. In analogy to the related experiment\cite{Izmalkov08} we plot
in Fig.~\ref{a_vs_fa_and_fb} the tank phase shift, when the tank is coupled
to the two flux qubits with the set of parameters (ii) and for the driving
frequency and amplitude: $\omega /2\pi =17.625$\ GHz and $f_{\mathrm{ac}%
}=10^{-3}$. The resonances are visualized as the ridge-trough lines, as
described above. The wide dip around $f_{b}\sim 0$ is due to the ground
state curvature. Solid lines are the energy contour lines which show where
the photon energy $\hbar \omega $ matches the respective levels: ground and
first excited (black), ground and second excited (white), and second excited
and upper (third excited) levels (orange/gray).

An interesting situation arises when one photon excites the system to an
intermediate level (to the second excited level in the figure), then, due to
the non-zero level population, another photon can excite the system to a
higher level (to the third excited level in the figure). This is shown with
the white circle in Fig.~\ref{a_vs_fa_and_fb}, which marks the region where
the signal is increased. The grey circle shows that such increase may be not
visible, when the former excitation (to the second excited level) creates
such population (close to $1/2$) that the tank signal is maximal.

\subsection{Multiphoton excitations}

\begin{figure}[h]
\includegraphics[width=8cm]{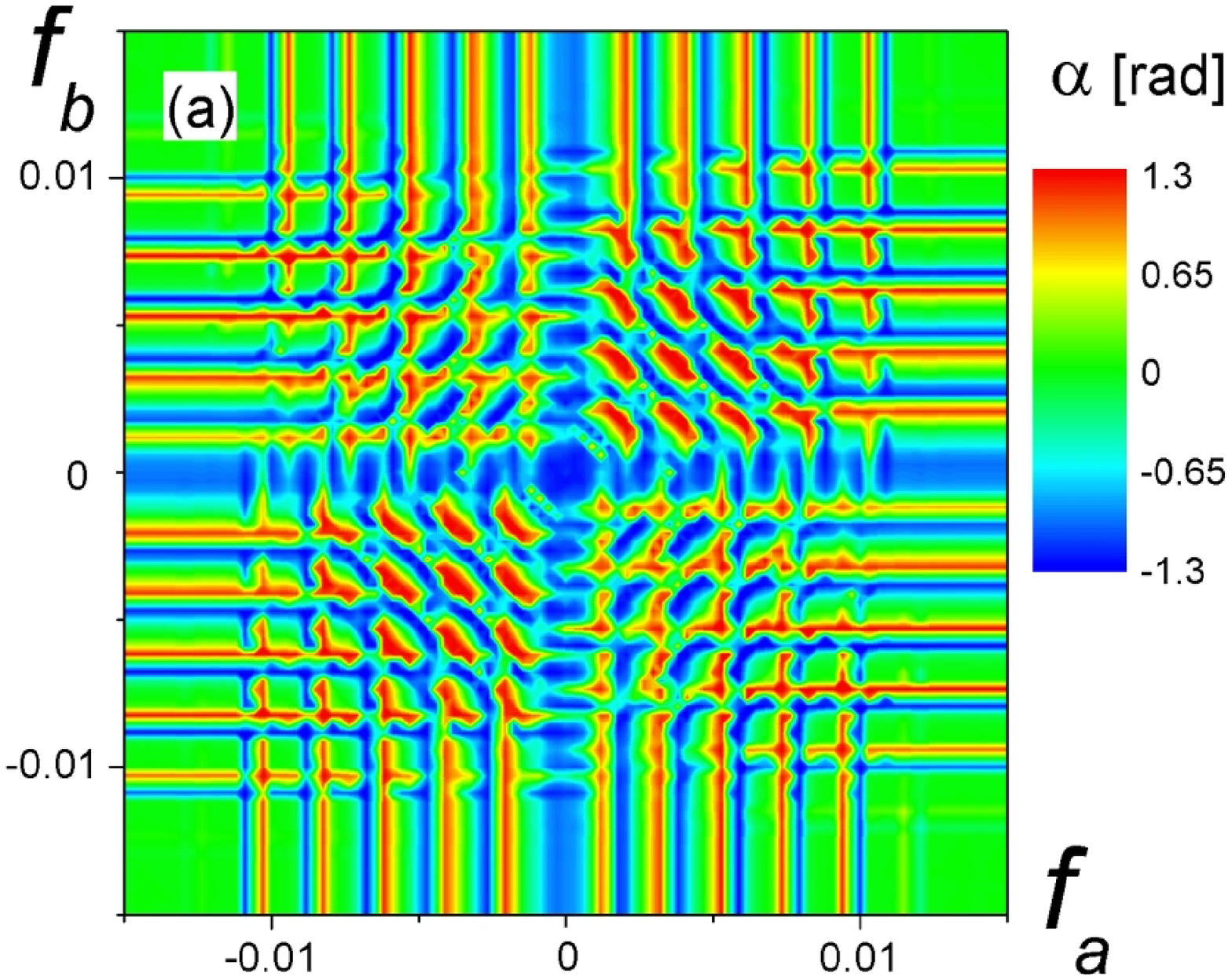} \includegraphics[width=8cm]{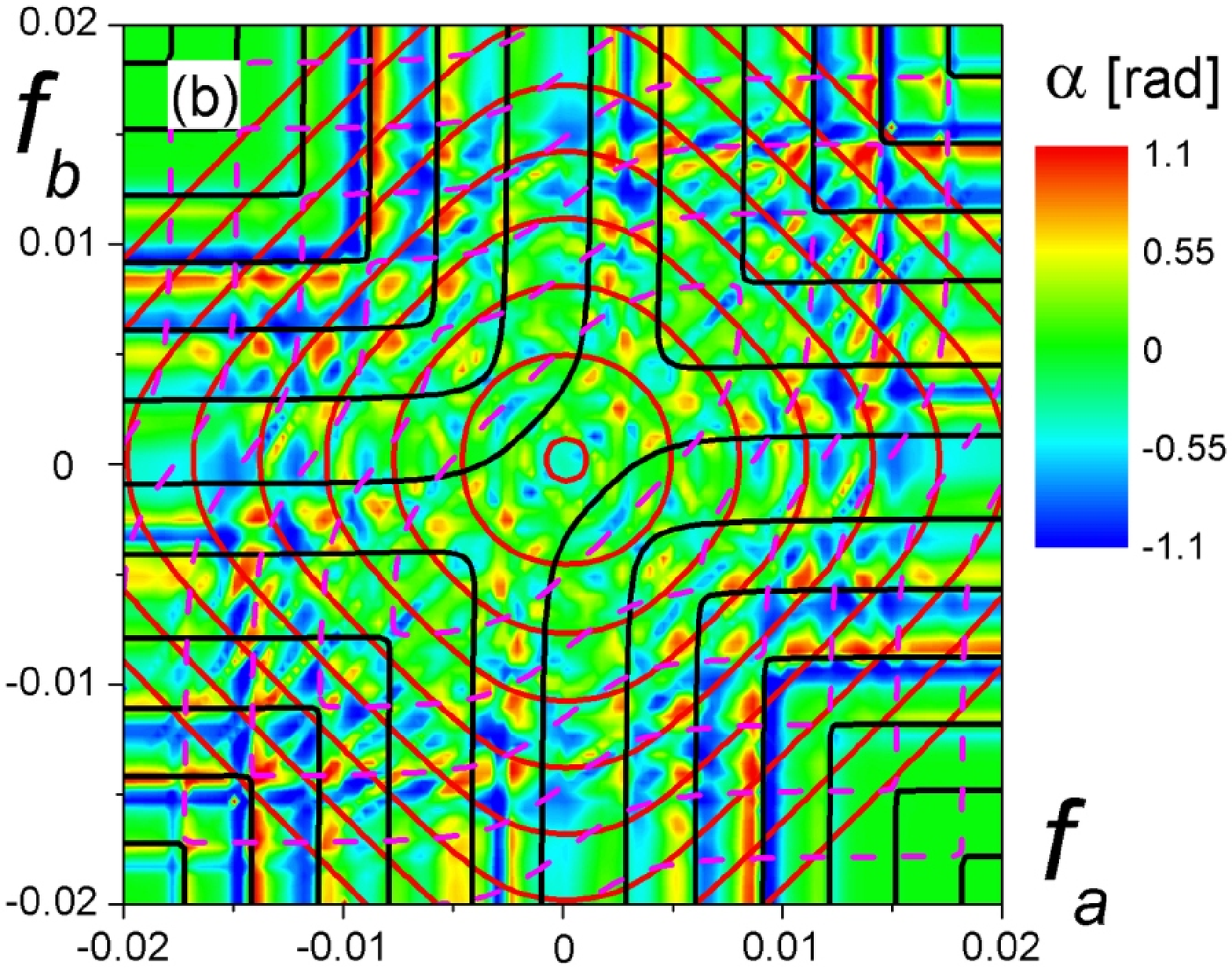}
\caption{(Color online). Multiphoton resonances in coupled qubits:
dependence of the tank phase shift on the flux detuning in qubits $a$ and $b$%
, $f_{a}$ and $f_{b}$.}
\label{multi4coupled}
\end{figure}

When the qubits are strongly driven, excitations due to multiphoton
processes become possible. In this case resonances appear where the energy
level difference is a multiple of the photon energy $\hbar\omega$. The
multiphoton fringes are shown in Fig.~\ref{multi4coupled} for the following
driving and qubit parameters: (a) $\omega/2\pi=4$\ GHz, $f_{\mathrm{ac}%
}=5\cdot10^{-3}$, and qubit parameters (i); (b) $\omega/2\pi=4.15$\ GHz and $%
f_{\mathrm{ac}}=7\cdot10^{-3}$, and qubit parameters from (ii) with equal
parameters (those for qubit $a$ are taken to be the same as for qubit $b$: $%
\Delta_{a}=\Delta_{b}=3.5$, etc.).

The lines in Fig. \ref{multi4coupled}(b) are the energy contour lines. They
show that the multiphoton resonances are mostly due to the excitation to the
first excited level (black lines) with the interruptions in ridge-trough
resonant structures (change of the signal), where higher levels are matched
with the multiple photon energy; possible excitations to the second excited
level are shown with the magenta (dash gray) line and to the upper level
with the red (solid dark gray) line.

\section{Relation to experiment}

As mentioned above, experimental results related to Figs. 2, 3b, and 5 were
already reported in Refs.~\onlinecite{Grajcar04} and \onlinecite{Izmalkov08}%
. The presented theory gives good agreement with those experiments. Here we
present some additional data obtained by making use of the same experimental
procedure as in Ref.~\onlinecite{Izmalkov08}. In these experiments the tank
circuit was realized as a parallel connection of a superconducting niobium
coil and commercial ceramic capacitor (see Fig.~1). The superconducting
persistent current qubits, three Josephson junctions closed into a
superconducting ring,\cite{Mooij} were deposited by an aluminum shadow
evaporation technique in the center of the superconducting coil. The voltage
across the tank circuit $V=v\exp (i\alpha )$ was amplified by a cryogenic
amplifier and measured by an $rf$ lock-in amplifier.

\begin{figure}[h]
\includegraphics[width=8cm]{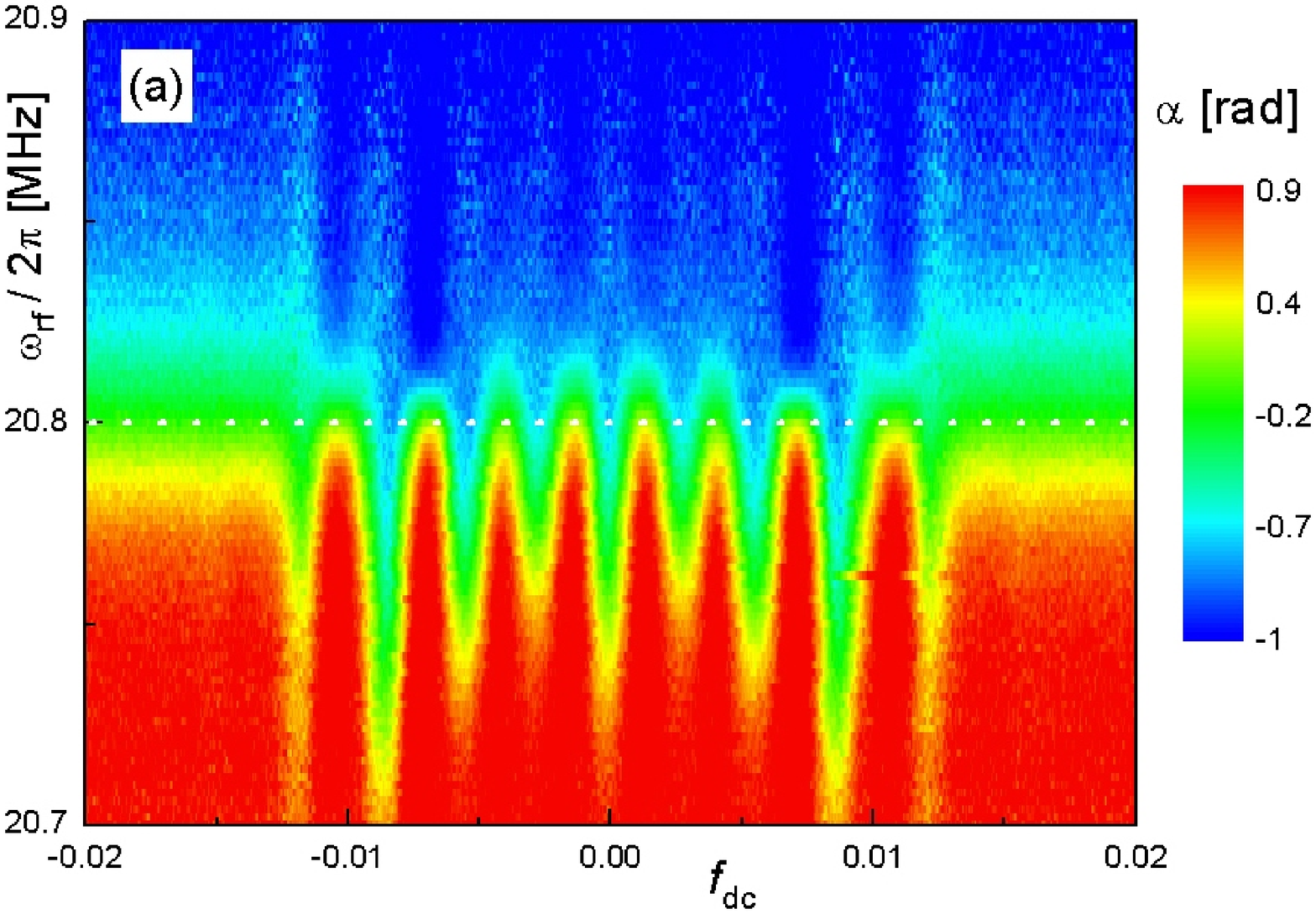} \includegraphics[width=8cm]{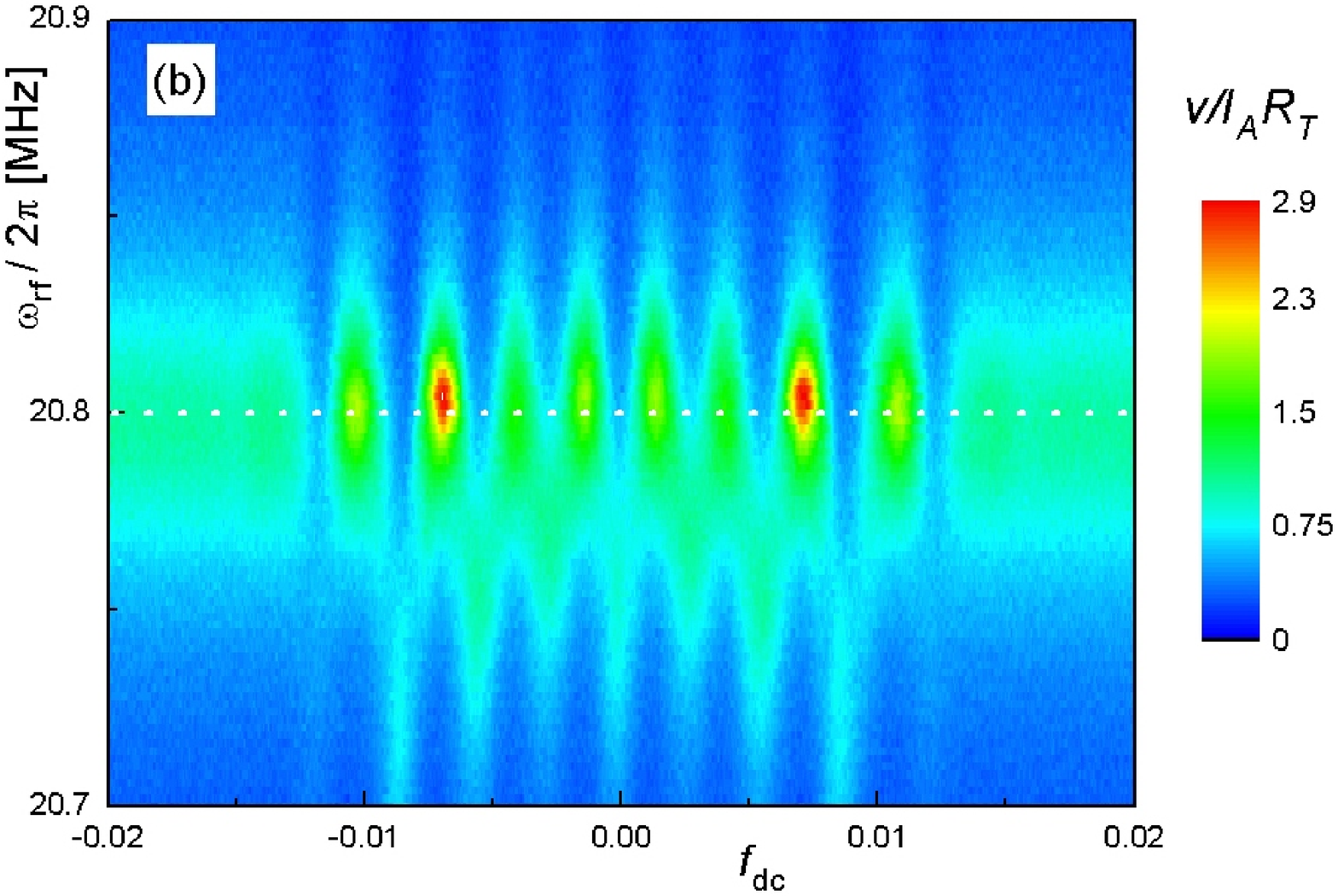}
\caption{(Color online). Experimental demonstration of multi-photon
excitations in a single flux qubit. (a) Dependence of the phase shift $%
\protect\alpha $ and (b) the amplitude $v$ on the bias current frequency $%
\protect\omega _{\mathrm{rf}}$ and the flux detuning $f_{\mathrm{dc}}$.
Recorded for a microwave excitation with $\protect\omega /2\protect\pi =4.15$
GHz and amplitude $f_{\mathrm{ac}}=4.5\cdot 10^{-3}$.}
\label{fig9}
\end{figure}

\begin{figure}[h]
\includegraphics[width=8cm]{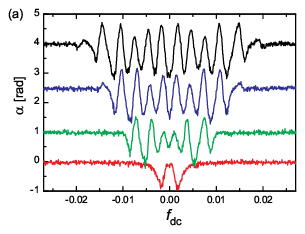} %
\includegraphics[width=8cm]{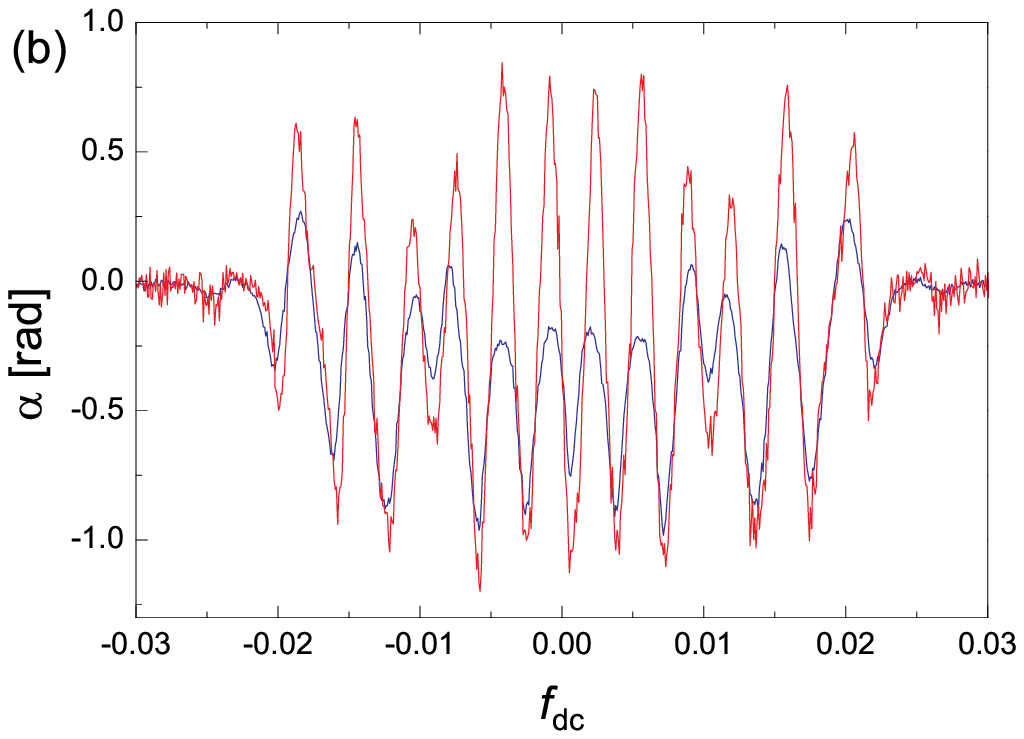}
\caption{(Color online). Experimental dependence of the phase shift $\protect%
\alpha $ on the bias flux $f_{\mathrm{dc}}$: (a) multiphoton resonances for
driving amplitudes $f_{\mathrm{ac}}/10^{-3}=0.5$, $2.5$, $4.5$, and $6.5$
from top to bottom; (b) influence of the bias current (red line $I_{A}M/\Phi
_{0}=2\cdot 10^{-6}$ and blue line - $I_{A}M/\Phi _{0}=8\cdot 10^{-6}$) for $%
\protect\omega /2\protect\pi =4.15$ GHz.}
\label{Fig10}
\end{figure}

In Fig. \ref{fig9} multi-photon resonances for a single qubit are shown in
dependence of both the DC flux $f_{\mathrm{dc}}$ and the frequency of the
driving current $\omega _{\mathrm{rf}}$. This data should be compared with
the theoretical result presented in Fig \ref{multiph_1qb}. Additional
multi-photon lines, similar to the theoretical results of Fig. \ref%
{spectroscopy}a, are shown in Fig. \ref{Fig10}a. Those results were recorded
at the resonant frequency only. Finally we studied the impact of finite bias
current on the tank's response as shown in Fig. \ref{Fig10}b which should be
compared with Fig. \ref{impact}. Also in this more detailed comparison with
the experiments we find a good agreement with the theory presented in this
work. Experimental results for the multi-photon resonances in coupled qubits
will be published elsewhere.

\section{Conclusion}

The application of the impedance measurement technique for monitoring
resonant excitations in flux qubits was studied. It was shown that
excitation of the qubits change their effective inductance. This results in
a changed effective inductance and resistance of the tank circuit. Thus, the
observable quantities -- the amplitude and phase shift of the tank voltage
-- reflect the changes of the effective inductances of the qubits.
Transparent expressions were derived in two limiting cases, one where the
dynamics of the tank is slow relatively to the time scale of the qubits and
another where both timescales are of the same order. In the first case the
changes in the tank's resistance are negligible and the phase shift is
directly related to the effective inductance. It has also been demonstrated
that in the latter case both the effective inductance and resistance of the
tank are defined by the inductance of the qubit. Our theoretical analysis
describes the state of the system of qubits in terms of the tank's effective
inductance and resistance. This allowed us to describe the experimental
results on multi-photon resonant excitation of single and coupled qubits.

The impact of the finite bias current and of the temperature was described
as well. In particular it was shown that the thermal excitation of the qubit
to the upper level results in the widening of the dip visible in the phase
shift as was observed previously in experiment\cite{Grajcar04}.

The resonant excitation of qubits can be used not only for controlling their
state, but also to determine their parameters. The developed theory has been
used to reproduce the experimental results presented in Ref.~%
\onlinecite{Izmalkov08} for single qubits and pairs of coupled flux qubits.
In addition to single-photon (\textquotedblleft
spectroscopic\textquotedblright ) resonances we have shown the appearance of
the multi-photon excitations in flux qubits coupled to the tank circuit.

\begin{acknowledgments}
We thank Ya.S. Greenberg and E.A. Ivanchenko for stimulating discussions.
This work was supported by the EU through the EuroSQIP project and by the
National Academy of Sciences of Ukraine under project \textquotedblleft
Nano\textquotedblright\ 2/07-H. S.N.S. acknowledges the financial support of
INTAS under YS Fellowship Grant No. 05-109-4479 and the hospitality of the
IPHT (Jena). M.G. was partially supported by the Slovak Scientific Grant
Agency Grant No. 1/0096/08, the Slovak Research and Development Agency under
the contract No. APVV-0432-07 and No. VCCE-005881-07, and Center of
Excellence of the Slovak Academy of Sciences (CENG).
\end{acknowledgments}

\end{document}